\documentclass[%
reprint,
superscriptaddress,
amsmath,amssymb,
aps,
prb,
floatfix
]{revtex4-2}

\usepackage{multirow}
\usepackage{graphicx}
\usepackage{dcolumn}
\usepackage{bm}
\usepackage[version=3]{mhchem}
\usepackage{nicefrac}

\usepackage{color}
\usepackage{xcolor}
\usepackage{dcolumn}
\usepackage{amssymb}
\usepackage{url}
\usepackage{hyperref}
\usepackage{xfrac}
\usepackage{tabularx}
\usepackage{xspace}
\usepackage{amsmath}
\usepackage[normalem]{ulem}
\usepackage{mathtools}
\usepackage{physics}
\usepackage{dsfont}
\usepackage{float}
\usepackage{braket}
\usepackage{placeins}
\usepackage{footmisc}

\def\be{\begin{equation}}
\def\ee{\end{equation}}
\def\bea{\begin{eqnarray}}
\def\eea{\end{eqnarray}}
\def\ba{\begin{array}}
\def\ea{\end{array}}

\begin{document}
\title{Landscape of incompressible crystals of hard-core bosons on the square-kagome lattice}
\author{Pratyay Ghosh}
\email{pratyay.ghosh@epfl.ch}
\affiliation{Institute of Physics, Ecole Polytechnique Fédérale de Lausanne (EPFL), CH-1015 Lausanne, Switzerland}

\begin{abstract}
We investigate a hard-core boson model on the square-kagome lattice using hierarchical mean-field theory beyond the conventional unit-cell description. We show that the conventional description of the square-kagome lattice based on its smallest unit cell does not fully capture the hierarchy of incompressible states supported by the lattice, but only captures the two compact-localized-state-based phases at densities $5/6$ and $2/3$. Within our approach, we reproduce these previously established phases and uncover additional incompressible states enabled by enlarging the variational cluster. Among these, the $\rho=3/4$ phase is found to be particularly robust, which, through the Matsubara-Matsuda mapping, corresponds to a half-magnetization plateau in the spin-$1/2$ XXZ model. We further apply our approach to two experimentally relevant square-kagome compounds using exchange parameters obtained from first-principles calculations. The calculated magnetization processes are in good agreement with available experimental results and predict additional plateau structures in these materials.
\end{abstract}

\maketitle

\section{Introduction}
The competition between kinetic energy and interactions plays a central role in the emergence of correlated quantum phases. Particularly interesting behavior arises in systems with nearly flat or perfectly flat bands, where the suppression of kinetic energy enhances the effects of interactions. In lattice systems, flat bands can emerge naturally from destructive quantum interference imposed by lattice geometry and symmetry. A characteristic feature of such flat-band systems is the existence of compact localized states (CLSs), exact eigenstates whose wave functions are confined to a finite region of the lattice. Unlike Anderson localization, CLSs arise in perfectly periodic systems and have been identified in a variety of frustrated hopping models, quantum spin systems, and interacting many-body systems~\cite{sutherland_localization_1986,bergman_band_2008,schulenburg_macroscopic_2002,richter_exact_2004,derzhko_universal_2006,derzhko_low-temperature_2010,Chen2023,ara_flat_2025}.

In quantum magnets, CLSs appear as localized magnon excitations, where flat-band single-spin-flip states become exact many-body eigenstates~\cite{richter_localized-magnon_2005,richter_exact_2004}. These localized magnons can generate macroscopic degeneracies and magnetization plateaus near the saturation field~\cite{schulenburg_macroscopic_2002,derzhko_low-temperature_2010,nakano_magnetization_2015,hasegawa_metamagnetic_2018}. More generally, frustration can stabilize localized structures also in ground states, such as exact product dimer and cluster states arising from competing exchange interactions~\cite{Majumdar1969,SriramShastry1981,Gelfand1991,Capponi2017,Ghosh2022,Ghosh2023}. 

\begin{figure*}[t]
    \includegraphics[width=0.95\textwidth]{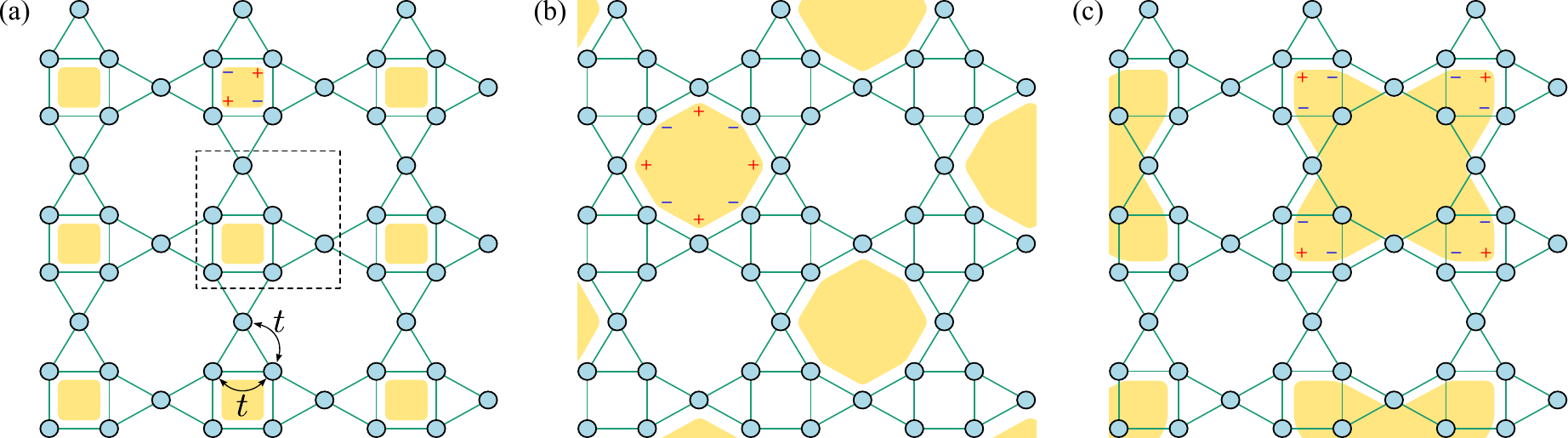}
    \caption{
    Compact localized states and approximate localized structures on the square-kagome lattice. 
    (a) Square-kagome lattice with nearest-neighbor hopping (or exchange) amplitude $t$. The dashed rectangle denotes the six-site unit cell. The yellow shaded squares indicate the elementary square trapping cells hosting compact localized states (CLS). The sign structure of the elementary square CLS wave function is indicated by red ($+$) and blue ($-$) symbols, representing the relative phases of the amplitudes.
    (b) Elementary octagon CLS. The alternating sign structure leads to destructive interference that confines the magnon within the octagonal plaquette.
    (c) Localized state on a 20-site cluster consisting of one octagon surrounded by four squares. Unlike the elementary square and octagon states, this state does not satisfy the destructive interference condition exactly and therefore is not a strict CLS, but exhibits an approximately localized amplitude structure.
    } \label{fig-lattice}
\end{figure*}

One natural setting where these ideas come together is the square-kagome lattice [see Fig.~\ref{fig-lattice}(a)], which comes with an exactly flat single-particle band generated by destructive interference~\cite{richter_localized-magnon_2005,schnack_exact_2006}. The antiferromagnetic Heisenebrg model on the square-kagome lattice (also known as the squagome or shuriken lattice in the literature) has been extensively studied in the context of frustrated quantum magnetism~\cite{siddharthan_square_2001,Rousochatzakis2013,derzhko_square-kagome_2014,lugan_topological_2019,richter_thermodynamics_2022,richter_magnetism_2023,astrakhantsev_pinwheel_2021,jahromi_quantum_2026}, where localized magnon states associated with this flat-band structure have been identified and provide a microscopic understanding of characteristic magnetization plateaus~\cite{Richter2009-vv,hasegawa_metamagnetic_2018,schmoll_tensor_2023,nakano_magnetization_2015}. Previous descriptions of localized states on the square-kagome lattice have primarily been based on taking the six-site unit cell as the fundamental cluster. This approach identifies two magnetization plateaus at $m/m_s=2/3$ and $1/3$ ($m_s$ is the saturation magnetization), respectively, in the spin-$1/2$ antiferromagnetic Heisenberg model on the square-kagome lattice, originating from localized magnon states~\cite{Richter2009-vv,nakano_magnetization_2015,richter_magnetism_2023,schmoll_tensor_2023}. 
While the six-site unit cell is the smallest natural choice, it is not the only possible cluster construction. Enlarging the cluster allows one to explore a broader set of configurations, including additional compact localized states and highly localized states that no longer satisfy the strict criteria of a CLS~\cite{ghosh2025simplexcrystalgroundstate}. This raises the possibility that larger clusters may reveal a richer set of incompressible states than those accessible within the six-site description.

To explore this question, we study interacting hard-core bosons on the square-kagome lattice (SKL) given by the Hamiltonian
\begin{equation}
\hat{H} =
t\sum_{\langle ij\rangle}
\left(
a_i^\dagger a_j+\mathrm{H.c.}
\right)
+
V\sum_{\langle ij\rangle}
n_i n_j
-
\mu\sum_i n_i ,
\end{equation}
where $t$ denotes the nearest-neighbor hopping amplitude, $V$ the nearest-neighbor repulsive interaction strength, and $\mu$ the chemical potential controlling the particle density. In this work, we consider only the positive-hopping case, $t>0$. The operators $a_i^\dagger$ ($a_i$) create (annihilate) hard-core bosons on site $i$, satisfying the constraint
$n_i=a_i^\dagger a_i\in\{0,1\}$. The hard-core boson model is directly related to quantum magnets through the Matsubara-Matsuda mapping~\cite{matsubara_lattice_1956},
\begin{equation}\label{eq:MM}
S_i^z=n_i-\frac12,\qquad
S_i^+=a_i^\dagger,\qquad
S_i^-=a_i,
\end{equation}
which relates hard-core bosons to spin-$1/2$ degrees of freedom. Within this correspondence, bosonic particles map onto spin flips, while the fully polarized state corresponds to the bosonic vacuum. Therefore, incompressible phases of the bosonic model directly translate into magnetic states of the corresponding spin system.

In this work, we systematically investigate localized and incompressible states of the hard-core boson model on the square-kagome lattice by progressively increasing the cluster size beyond the conventional six-site unit cell. We employ a cluster mean-field approach in a hierarchical manner, where larger clusters provide an enlarged variational space for exploring localized configurations and incompressible phases. The enlarged clusters allow access to additional compact localized states that are not captured within the conventional single-unit-cell description. The increased cluster size reveals additional incompressible states that do not strictly satisfy the compact localization condition of CLSs. Finally, we apply our approach to two experimentally relevant square-kagome compounds, namely \ce{KCu6AlBiO4(SO4)5Cl} and \ce{Na6Cu7BiO4(PO4)4Cl3}~\cite{fujihala_gapless_2020,Yakubovich2021,liu_low-temperature_2022},  obtaining magnetization curves in good agreement with available experimental results and predicting additional plateau structures.

\section{Compact Localized States in SKL}\label{sec:CLS}

Let us first briefly review the construction of localized-magnon states. Following the prescription of Richter \emph{et al.}~\cite{richter_exact_2004,richter_localized-magnon_2005}, one starts from single-magnon excitations above the fully polarized (FP) state. For a magnetic field $h>h_s$, the lowest excitation above the FP state is a single magnon, which is generally dispersive. However, on certain frustrated lattices, such as the square-kagome lattice (SKL), destructive interference can suppress magnon propagation and lead to strictly localized eigenstates.

Consider a single-magnon wavefunction confined within a finite region $\mathcal{R}$,
\begin{equation}
\ket{\Phi}_{\mathcal{R}}
\sim
\sum_{i\in\mathcal{R}} c_i S_i^-\ket{\mathrm{FP}},
\label{eq:magnon}
\end{equation}
where $c_i$ denotes the amplitude of the magnon on site $i$. For this state to remain localized, it must be an eigenstate of the Hamiltonian restricted to the region $\mathcal{R}$, while the amplitudes leaking into the surrounding region $\bar{\mathcal{R}}$ must vanish. Since the magnon can leave $\mathcal{R}$ through the exchange couplings connecting the two regions, destructive interference between these hopping processes is required. This condition is expressed as
\begin{equation}
\sum_{i\in\mathcal{R}} J_{ij}c_i=0,
\qquad
\forall j\in\bar{\mathcal{R}} .
\end{equation}
This condition is necessary for the existence of a CLS.

On the SKL, this condition is satisfied for the simplest choice of a magnon localized on an elementary square plaquette, with an alternating sign structure of the coefficients $c_i$, as shown in Fig.~\ref{fig-lattice}(a). By placing such square CLSs on every square plaquette, one can construct an exact product state of localized magnons corresponding to the magnetization $m/m_s=2/3$. This state becomes energetically favorable with respect to the fully polarized state below the saturation field. A second localized-magnon crystal can be constructed by considering two-magnon states confined within each square plaquette, which gives rise to the incompressible phase at $m/m_s=1/3$.

\begin{figure*}[t]
    \includegraphics[width=0.95\textwidth]{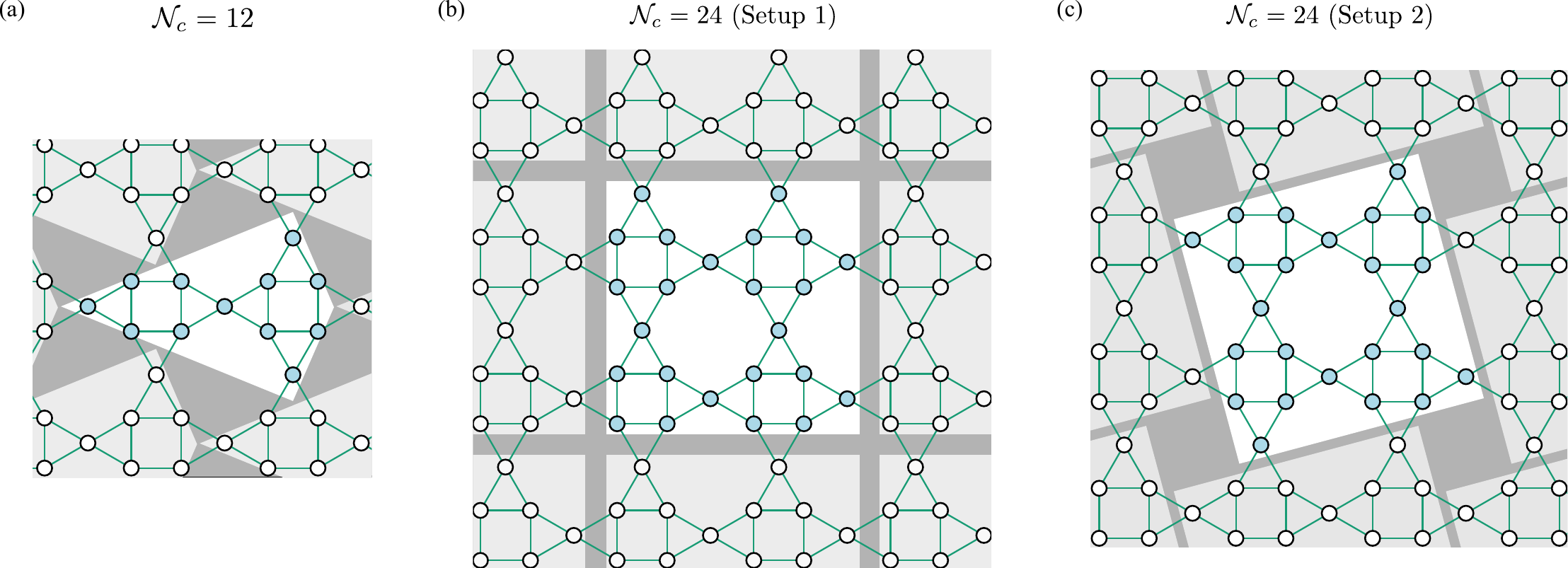}
    \caption{Cluster geometries used in the hierarchical mean-field calculations.
(a) The 12-site cluster, which represents the smallest cluster considered in this work. 
(b) The 24-site cluster (Setup 1), chosen to accommodate the different localized structures discussed in the text.
(c) The alternative 24-site cluster (Setup 2), which contains the same number of sites as Setup 1 but preserves the fourfold rotational symmetry around the centers of the octagonal plaquettes.} \label{fig-cluster}
\end{figure*}

Beyond the square-plaquette localized states, the SKL also permits other localized-magnon constructions. For example, a magnon can be confined to an elementary octagonal plaquette, as shown in Fig.~\ref{fig-lattice}(b). These octagonal localized states provide an additional type of localized building block and can also be arranged into crystalline patterns of localized magnons. Larger localized structures can also be considered by taking clusters containing one octagon surrounded by four squares [Fig.~\ref{fig-lattice}(c)]. The Hamiltonian restricted to this 20-site region admits an eigenstate with an approximately alternating amplitude structure: the amplitudes on the corner sites and their neighboring sites have nearly equal magnitude and opposite sign. As a result, the destructive interference condition is approximately satisfied, and the state exhibits a strongly localized character. Unlike the square and octagonal cases, this state is not an exact CLS, since a finite leakage amplitude remains outside the cluster. Nevertheless, such larger-cluster localized states provide additional configurations that can be stabilized within an enlarged variational space and may lead to incompressible phases beyond those associated with exact CLSs.

\section{Methods}\label{sec:methods}
\subsection{Hierarchical mean-field theory}
To investigate the ground-state properties, we employ a cluster mean-field approach, which we use in a hierarchical manner by systematically increasing the cluster size [see examples in Fig.~\ref{fig-cluster}]. We refer to this approach as hierarchical mean-field theory (HMFT).
The method treats short-range quantum correlations exactly within finite clusters while approximating correlations between clusters at the mean-field level.
The method has been successfully applied to a variety of strongly correlated lattice models and is particularly well suited for frustrated systems, where local quantum fluctuations play a crucial role~\cite{HMFT1,HMFT2,HMFT3,HMFT4,HMFT5,huerga_staircase_2016}.

Within HMFT, the lattice is partitioned into identical clusters that tile the infinite lattice without overlap. The Hamiltonian is correspondingly decomposed into intra-cluster and inter-cluster contributions,
\begin{equation}
\hat{H}=\sum_c \hat{H}_c+\sum_{\langle c,c'\rangle}\hat{H}_{cc'},
\end{equation}
where $\hat{H}_c$ contains all hopping and interaction terms entirely within cluster $c$, while $\hat{H}_{cc'}$ couples neighboring clusters through bonds crossing the cluster boundary.

The many-body wave function is approximated by a product of cluster wave functions,
\begin{equation}
|\Psi\rangle=\prod_c |\psi_c\rangle,
\end{equation}
which preserves the full quantum correlations inside each cluster while treating inter-cluster correlations in a self-consistent mean-field approximation. The boundary hopping terms are decoupled according to
\begin{equation}
a_i^\dagger a_j
\approx
\langle a_i^\dagger\rangle a_j
+a_i^\dagger\langle a_j\rangle
-\langle a_i^\dagger\rangle\langle a_j\rangle,
\end{equation}
while density-density interactions are treated as
\begin{equation}
n_i n_j
\approx
\langle n_i\rangle n_j
+n_i\langle n_j\rangle
-\langle n_i\rangle\langle n_j\rangle.
\end{equation}
The resulting effective cluster Hamiltonian depends on the expectation values of boundary operators, which are determined self-consistently.

For a given set of mean fields, the effective cluster Hamiltonian is solved exactly by numerical diagonalization within the hard-core boson Hilbert space. The expectation values of the boundary operators are then updated from the resulting ground state and the procedure is iterated until convergence. Throughout this work, convergence is assumed when successive iterations change all mean fields by less than a numerical tolerance of $10^{-6}$.

\subsection{Choice of cluster geometry}
An important aspect of the HMFT approach is that the choice of cluster defines the variational space accessible to the calculation. Since the many-body wave function is constructed as a product of cluster wave functions, only spatial structures and ordering patterns compatible with the chosen cluster geometry can be represented. By systematically enlarging the cluster, HMFT provides a hierarchy of approximations with an increasingly rich variational space, allowing competing localized states to be explored on an equal footing. 

The SKL supports several localized configurations, including compact localized states and states with approximately localized character, as illustrated in Fig.~\ref{fig-lattice}. To investigate these possibilities, we analyze the spatial extent and symmetry properties of these localized modes and then select cluster geometries capable of accommodating them. The three clusters considered in this work are shown in Fig.~\ref{fig-cluster}. The first cluster contains 12 sites and represents the smallest cluster considered in our calculations [see Fig.~\ref{fig-cluster} (a)]. While it can accommodate the CLS shown in Fig.~\ref{fig-lattice}(a), its geometry is insufficient to represent the other localized configurations shown in Fig.~\ref{fig-lattice}. We therefore consider two additional 24-site clusters. The first cluster, referred to as Setup 1 [see Fig.~\ref{fig-cluster} (b)], is the smallest cluster that can simultaneously accommodate all candidate localized structures shown in Fig.~\ref{fig-lattice}. The second cluster, Setup 2 [see Fig.~\ref{fig-cluster} (c)], contains the same number of sites but is chosen to preserve the fourfold rotational symmetry around the centers of the octagonal plaquettes. Together, these three cluster geometries allow us to assess the robustness of the incompressible phases with respect to the accessible variational space and the symmetry of the cluster construction.



\section{Connection to the XXZ spin model}

Through the Matsubara--Matsuda transformation introduced in Eq.~\eqref{eq:MM}, the hard-core boson Hamiltonian can be mapped exactly onto a spin-$1/2$ XXZ model in an external magnetic field. In terms of spin operators, the Hamiltonian becomes
\begin{equation}
\hat{H}
=
2t\sum_{\langle ij\rangle}
\left(
S_i^xS_j^x+S_i^yS_j^y
\right)
+
V\sum_{\langle ij\rangle}S_i^zS_j^z
-
(\mu-2V)\sum_i S_i^z .
\end{equation}
Thus, the hopping amplitude $t$ controls the transverse exchange interaction, while the nearest-neighbor repulsion $V$ determines the longitudinal exchange coupling. The chemical potential in the bosonic description corresponds to a magnetic field along the $z$ direction.

This correspondence allows the results obtained in the hard-core boson language to be directly interpreted in terms of magnetic phases of the spin-$1/2$ XXZ model. By varying the ratio between $t$ and $V$, the model continuously interpolates between different exchange anisotropies, including the Ising limit ($t\rightarrow0$), the XY limit ($V\rightarrow0$), and the isotropic Heisenberg point ($V=2t$). For convenience, we introduce a dimensionless parametrization of the coupling ratios~\cite{huerga_staircase_2016},
\begin{equation}
t'=\frac{t}{\sqrt{t^2+V^2}},
\qquad
\mu'=\frac{\mu-2V}{2(t+V)},
\end{equation}
which is used to represent the phase diagram throughout the paper. The corresponding spin-model limits are obtained at $t'=0$ for the Ising limit, $t'=1/\sqrt{5}$ for the isotropic Heisenberg point, and $t'=1$ for the XY limit.

\section{Results: Ground-state phase diagram}
\begin{figure*}[t]
    \includegraphics[width=0.9\textwidth]{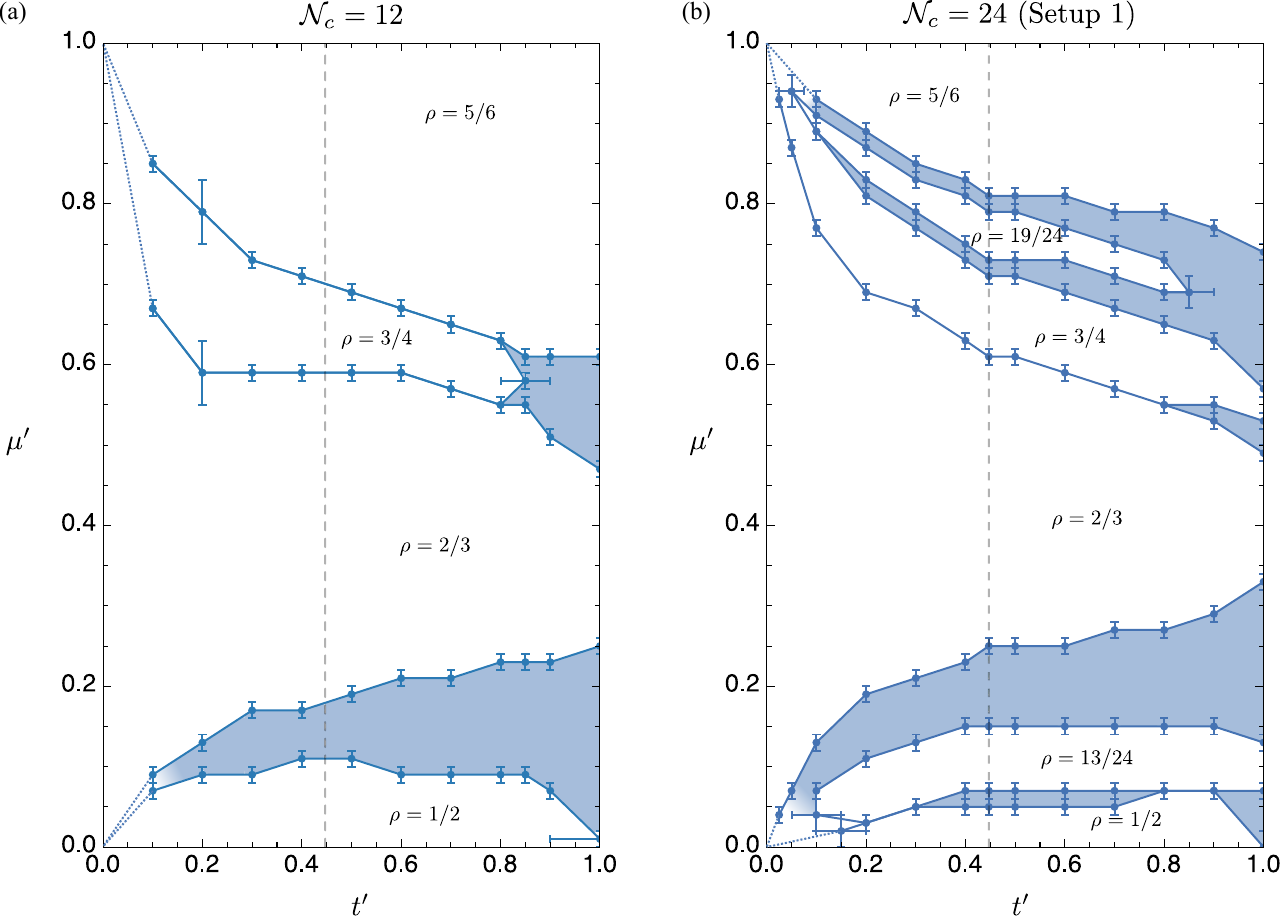}
    \caption{
Ground-state phase diagrams obtained from cluster mean-field theory.
(a) Phase diagram obtained using the 12-site cluster. The blue regions denote compressible phases with continuously varying density, while the other regions correspond to incompressible phases. The incompressible plateaus appear at $\rho=1/2$, $2/3$, $3/4$, and $5/6$.
(b) Phase diagram obtained using the 24-site cluster (Setup 1). The overall phase diagram is in qualitative agreement with the 12-site result, while the enlarged cluster stabilizes additional incompressible phases at $\rho=13/24$ and $\rho=19/24$. } \label{fig:QPD_1}
\end{figure*}

\begin{figure*}[t]
    \includegraphics[width=0.95\textwidth]{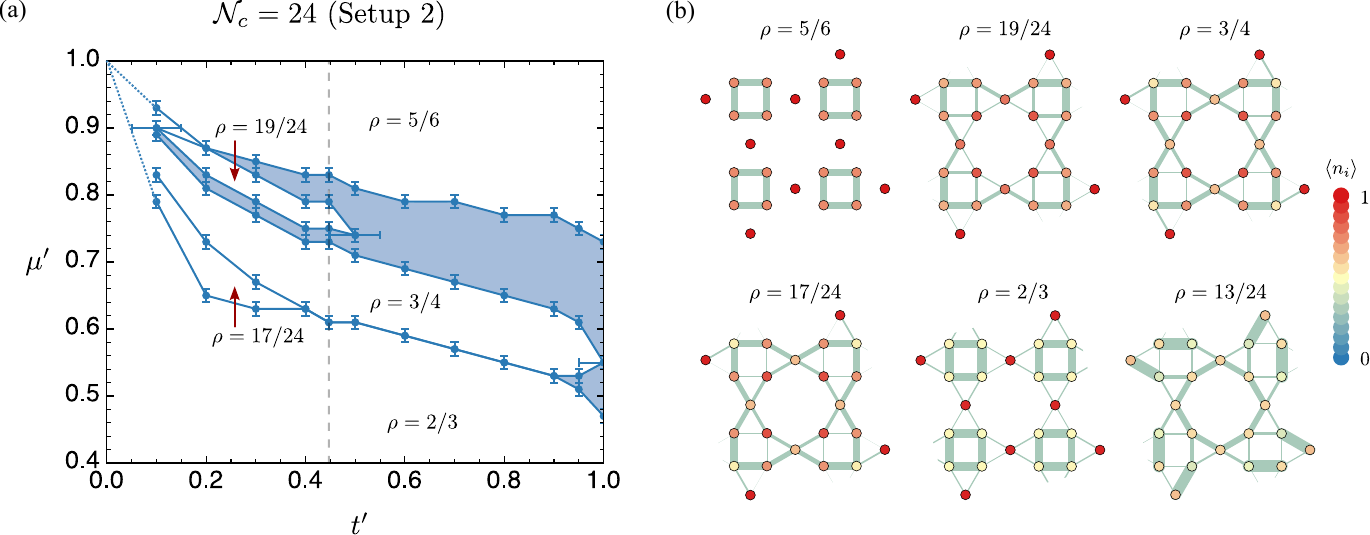}
    \caption{
    Comparison of the phase diagram obtained using the alternative 24-site cluster (Setup 2) and the corresponding local observables.
(a) Upper part of the phase diagram ($\mu'\geq 0.4$) calculated using Setup 2. The lower-density region is qualitatively identical to the result obtained from Setup 1 and is not shown. Compared with Setup 1, the $\rho=19/24$ incompressible phase becomes smaller, while an additional incompressible phase emerges at $\rho=17/24$, located between the $\rho=2/3$ and $\rho=3/4$ plateaus.
(b) Spatial distribution of the local density $n_i=\langle a_i^\dagger a_i\rangle$ and the hopping expectation value $B_{ij}=\langle a_i^\dagger a_j+\mathrm{H.c.}\rangle$ for the corresponding insulating states. The color scale represents the local density, while the magnitude of $B_{ij}$ is indicated by the thickness of the bonds.
}\label{fig:QPD_2}
\end{figure*}

The ground-state phase diagrams obtained from the HMFT calculations are presented in Figs.~\ref{fig:QPD_1} and \ref{fig:QPD_2}. Figure~\ref{fig:QPD_1}(a) shows the results obtained using the 12-site cluster, while Fig.~\ref{fig:QPD_1}(b) corresponds to the 24-site cluster (Setup 1). To assess the dependence of the results on the cluster geometry, we additionally consider the alternative 24-site cluster (Setup 2), whose phase diagram is shown in Fig.~\ref{fig:QPD_2}(a). 
Owing to the particle-hole symmetry of the hard-core boson Hamiltonian, the phase diagram is symmetric about half filling, $\rho=1/2$. Consequently, it is sufficient to discuss the phases for $\mu'\geq0$, while the corresponding phases for $\mu'\leq0$ follow directly from symmetry.

The 12-site calculation reproduces the insulating phases previously reported for the square-kagome lattice~\cite{Richter2009-vv,nakano_magnetization_2015,richter_magnetism_2023,schmoll_tensor_2023}. Besides the half-filled insulating phase at $\rho=1/2$, corresponding to the zero-magnetization plateau ($m/m_s=0$), robust incompressible plateaus are found at $\rho=2/3$ and $5/6$. These latter two phases correspond to the localized-magnon crystal magnetization plateaus at $m/m_s=1/3$ and $2/3$, respectively, in the spin-$1/2$ antiferromagnetic Heisenberg model on the square-kagome lattice. An additional incompressible phase also appears at $\rho=3/4$, corresponding to the $m/m_s=1/2$ plateau, demonstrating that the 12-site cluster already stabilizes an ordered phase beyond the conventional localized-magnon regime.

A qualitatively richer phase diagram emerges when the cluster size is increased to 24 sites. In addition to all phases present in the 12-site calculation, the larger cluster stabilizes new incompressible plateaus at $\rho=13/24$ and $19/24$, corresponding to magnetization plateaus at $m/m_s=1/12$ and $7/12$, respectively. To verify that these additional phases are intrinsic to the model rather than artifacts of a particular cluster construction, we repeated the calculations using the alternative 24-site cluster (Setup 2), which preserves the fourfold rotational symmetry about the centers of the octagonal plaquettes. The resulting phase diagram exhibits the same overall topology as that obtained from Setup 1, reproducing all previously identified incompressible phases while additionally stabilizing a plateau at $\rho=17/24$, corresponding to a magnetization plateau at $m/m_s=5/12$. These results suggest that the SKL hosts a richer hierarchy of incompressible phases than previously recognized.

Among the additional incompressible phases revealed by the enlarged clusters, the plateaus at $\rho=19/24$, $\rho=3/4$, and $\rho=17/24$ can be understood in terms of the larger 20-site approximate localized state discussed earlier. The corresponding density and hopping patterns depicted in Fig.~\ref{fig:QPD_2}(b) show that the state is strongly localized within the 20-site region, with only weak hopping processes extending beyond the cluster boundary. Thus, these phases do not originate from exact CLSs, but rather from highly localized configurations with an approximate compact structure.

While the incompressible phases at $\rho=13/24$, $\rho=17/24$, and $\rho=19/24$ become accessible only upon enlarging the cluster size, the $\rho=3/4$ incompressible phase is already present in both the 12-site and 24-site calculations. Its persistence across different cluster geometries indicates that this phase is a robust feature of the square-kagome lattice. We note that previous exact diagonalization studies in Refs.~\cite{nakano_magnetization_2015,richter_thermodynamics_2022} also reported signatures of the corresponding plateau at $\rho=3/4$ ($m/m_s=1/2$) for clusters containing 24, 36, and 48 sites. However, these features were not identified as a robust incompressible phase and were attributed to possible finite-size effects. The localized density and bond patterns obtained from our enlarged-cluster HMFT calculations instead suggest that the $\rho=3/4$ phase originates from a specific approximately localized configuration supported by the square-kagome lattice.

\section{Discussion about material realizations}
Having established the phase diagram of the ideal square-kagome model, we now apply our approach to two experimentally relevant compounds, namely \ce{KCu6AlBiO4(SO4)5Cl} and \ce{Na6Cu7BiO4(PO4)4Cl3}~\cite{fujihala_gapless_2020,Yakubovich2021,liu_low-temperature_2022}, whose magnetic interactions have been obtained from first-principles calculations~\cite{fujihala_gapless_2020,niggemann_quantum_2023}. Both materials are well described by spin-$1/2$ Heisenberg models, but depart from the ideal square-kagome geometry through additional bonds or exchange anisotropies. 

\subsection{\ce{KCu6AlBiO4(SO4)5Cl}}
\begin{figure*}[t]
    \includegraphics[width=0.95\textwidth]{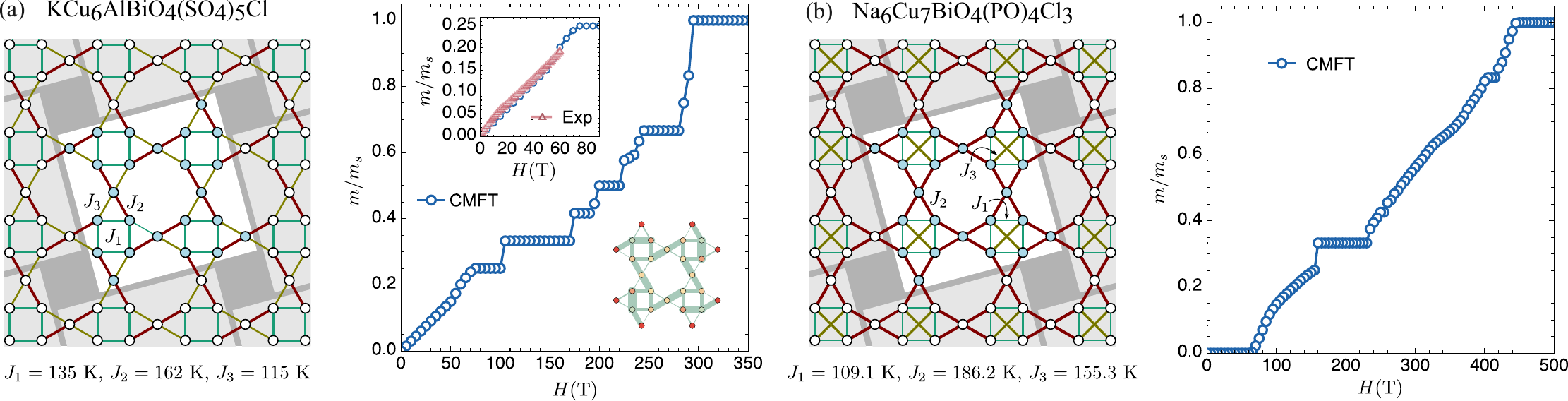}
    \caption{
    Magnetic properties of square-kagome compounds. 
    (a) Exchange network and calculated magnetization curve of \ce{KCu6AlBiO4(SO4)5Cl} using the 24-site Setup 2. The calculated results show good agreement with available experimental data, reproducing the observed $m/m_s=1/3$ and $2/3$ plateaus and predicting additional plateau structures, with the lowest-field plateau appearing at $m/m_s=1/4$. The inset shows the local density and hopping expectation values of the $m/m_s=1/4$ plateau.
    (b) Exchange network and calculated magnetization curve of \ce{Na6Cu7BiO4(PO)4Cl3}. The system exhibits plateaus at $m/m_s=1/3$ and $5/6$, with a shoulder near $m/m_s=2/3$.
    } \label{fig:compounds}
\end{figure*}

We first consider \ce{KCu6AlBiO4(SO4)5Cl}, whose exchange parameters are taken from Ref.~\cite{fujihala_gapless_2020}. The corresponding spin model is shown in Fig.~\ref{fig:compounds}(a). Unlike the ideal square-kagome lattice, the exchange interactions are anisotropic: the bonds forming the square plaquettes are characterized by an exchange coupling $J_1$, while the octagonal plaquettes are dimerized with alternating couplings $J_2$ and $J_3$. The anisotropy between $J_2$ and $J_3$ breaks the destructive interference condition required for an exact CLS on the ideal lattice; therefore, the corresponding plateau states are no longer localized magnon states.

The magnetization curve was computed using the 24-site Setup 2 introduced in Sec.~\ref{sec:methods}. The resulting magnetization process is shown in Fig.~\ref{fig:compounds}(a). We find a sequence of magnetization plateaus at
\[
m/m_s=\frac{1}{4},\,
\frac{1}{3},\,
\frac{5}{12},\,
\frac{1}{2},\,
\frac{2}{3}.
\]
A direct comparison with experiment reveals excellent agreement. Measurements performed up to magnetic fields of $60~\mathrm{T}$ show a continuous increase of the magnetization from zero field without any low-field plateau, in quantitative agreement with our calculations [see inset of Fig.~\ref{fig:compounds}(a)]. At higher magnetic fields, the experimental study further reports isolated measurements of magnetization plateaus at $m/m_s=1/3$ and $2/3$, located near $150~\mathrm{T}$ and $270~\mathrm{T}$, respectively. Both plateaus are reproduced by our calculations, providing strong support for the microscopic exchange model of Ref.~\cite{fujihala_gapless_2020}.

An important prediction of the present work is the existence of an additional magnetization plateau at $m/m_s=1/4$, which is expected to appear at a magnetic field of approximately $75~\mathrm{T}$. Since this field lies above the range explored in the initial low-field measurements but well below the experimentally observed $1/3$ plateau, it should be accessible to future high-field magnetization experiments.

\subsection{\ce{Na6Cu7BiO4(PO4)4Cl3}}
We next consider \ce{Na6Cu7BiO4(PO4)4Cl3} using the dominant exchange parameters reported in Ref.~\cite{niggemann_quantum_2023}. The corresponding exchange network is shown in Fig.~\ref{fig:compounds}(b). In addition to the nearest-neighbor couplings on the square-kagome lattice, this material possesses a strong diagonal exchange interaction $J_3$ across the elementary squares. Notably, this diagonal interaction is larger than the nearest-neighbor square interaction $J_1$, substantially modifying the magnetic ground state.

The calculated magnetization curve is presented in Fig.~\ref{fig:compounds}(b). We find that the magnetization remains zero up to approximately $60~\mathrm{T}$ before increasing continuously with field. At higher fields, well-defined magnetization plateaus appear at $m/m_s=1/3,$ and $5/6,$
while only a weak shoulder is observed near
$m/m_s=2/3$
rather than a fully developed plateau.

The presence of the strong diagonal exchange does not invalidate the compact localized-state construction described in Sec.~\ref{sec:CLS}; an exact CLS can still be constructed by the same destructive-interference mechanism. However, the modified exchange hierarchy shifts this state above other one-magnon excitations, so that it no longer constitutes the lowest-energy state below saturation. Consequently, the magnetization jump characteristic of the ideal square-kagome lattice disappears, while the highest-field plateau occurs at $m/m_s=5/6$.

\section{Conclusions}
In this work, we investigated the hard-core boson model on the SKL using HMFT with systematically increasing cluster sizes. Motivated by the observation that previous studies were largely restricted to the six-site unit cell, we explored how enlarging the variational cluster modifies the landscape of localized and incompressible states. We demonstrated that larger clusters provide access to a significantly richer set of insulating phases that remain hidden within conventional cluster descriptions.

The resulting ground-state phase diagram reveals a richer hierarchy of incompressible phases than previously recognized in the square-kagome lattice. We recover the previously reported incompressible states at $\rho=2/3$ and $\rho=5/6$~\cite{Richter2009-vv,nakano_magnetization_2015,richter_magnetism_2023,schmoll_tensor_2023}, while the enlarged clusters reveal additional incompressible phases at $\rho=13/24$, $\rho=17/24$, $\rho=19/24$, and $\rho=3/4$. Through the Matsubara-Matsuda mapping, these phases correspond directly to new magnetization plateaus in the spin-$1/2$ XXZ model. The phases at $\rho=13/24$, $\rho=17/24$, and $\rho=19/24$ emerge only when larger clusters are considered, highlighting the importance of extended correlations beyond the conventional six-site description. In contrast, the $\rho=3/4$ phase is already stabilized in both the 12-site and 24-site calculations, suggesting that it represents a particularly robust feature of the SKL. More generally, these results show that enlarging the cluster in HMFT can reveal incompressible phases that are inaccessible within smaller cluster descriptions. In contrast, comparing the results from different cluster geometries provides a way to assess their robustness.

Finally, we applied our approach to the experimentally relevant square-kagome compounds \ce{KCu6AlBiO4(SO4)5Cl} and \ce{Na6Cu7BiO4(PO4)4Cl3}~\cite{fujihala_gapless_2020,Yakubovich2021,liu_low-temperature_2022,niggemann_quantum_2023} using exchange parameters obtained from first-principles calculations. For \ce{KCu6AlBiO4(SO4)5Cl}, the calculated magnetization curve is in good agreement with available experimental measurements~\cite{fujihala_gapless_2020} and predicts an additional $1/4$ magnetization plateau that can be tested in future high-field experiments. For \ce{Na6Cu7BiO4(PO4)4Cl3}, we demonstrate that additional exchange interactions significantly modify the high-field magnetization process, leading to behavior that differs qualitatively from the ideal square-kagome antiferromagnetic Heisenberg model.

\textit{Acknowledgments:}
P.G.~acknowledges financial support from the Swiss National Funds.


%

\end{document}